\documentclass[conference]{IEEEtran}
\IEEEoverridecommandlockouts
\usepackage{amsmath,amssymb,amsfonts}
\usepackage{algorithmic}
\usepackage{graphicx}
\usepackage{textcomp}
\usepackage[dvipsnames]{xcolor}
\usepackage{personalcommand}
\usepackage[numbers]{natbib}
\usepackage{comment}
\usepackage{hyperref}
\usepackage{cleveref}

\def\BibTeX{{\rm B\kern-.05em{\sc i\kern-.025em b}\kern-.08em
    T\kern-.1667em\lower.7ex\hbox{E}\kern-.125emX}}
\begin{document}

\title{Quality Attributes Optimization of Software Architecture: Research Challenges and Directions}

\author{\IEEEauthorblockN{Daniele Di Pompeo}
\IEEEauthorblockA{\textit{University of L'Aquila} \\
L'Aquila, Italy \\
daniele.dipompeo@univaq.it}
\and
\IEEEauthorblockN{Michele Tucci}
\IEEEauthorblockA{\textit{Charles University} \\
Prague, Czech Republic \\
tucci@d3s.mff.cuni.cz}
}

\maketitle

\begin{abstract}
The estimation and improvement of quality attributes in software architectures is a challenging and time-consuming activity. On modern software applications, a model-based representation is crucial to face the complexity of such activity.
One main challenge is that the improvement of distinctive quality attributes may require contrasting refactoring actions on the architecture, for instance when looking for trade-off between performance and reliability (or other non-functional quality attributes). In such cases, multi-objective optimization can provide the designer with a more complete view on these trade-offs and, consequently, can lead to identify suitable refactoring actions that take into account independent or even competing objectives.


In this paper, we present open challenges and research directions to fill current gaps in the context of multi-objective software architecture optimization.

\end{abstract}


\section{Introduction}\label{sec:intro}

Different factors, such as the addition of new requirements, the adaption to new execution contexts, or the deterioration of non-functional attributes, can lead to software refactoring.
Identifying the best refactoring operations is challenging because there is a wide range of potential solutions and no automated assistance is currently available.
In this situation, search-based approaches have been widely used \citep{Mariani:2017jd,Ouni:2017db,Ramirez:2018uz,Ray:2014ip,Aleti:2013gp}.

Multi-objective optimization approaches, which are search-based, have lately been used to solve model refactoring optimization issues~\citep{CORTELLESSA2021106568,NI2021106565}.
Searching among design alternatives (for example, through architectural tactics) is a typical feature of multi-objective optimization methodologies used to solve model-based software restructuring challenges~\citep{Koziolek:2011cg,NI2021106565}.

The automated refactoring of software models plays an important role in optimizing software architectures, as it allows generating design alternatives while preserving the external behavior of its functionalities.
While being beneficial in finding such alternatives, the automated refactoring process can generate a considerable number of new solutions that are difficult for the designer to navigate.
%
As a result, choosing the best refactoring methods from such a huge set of options requires significant effort, which can be reduced by multi-objective algorithms.
However, in order to explore the solution space and produce an (almost) optimal Pareto frontier, multi-objective algorithms may require a significant amount of hardware resources (such as time and memory allocation).
Even when automated, finding and creating Pareto boundaries can frequently take many hours or even days. 
Therefore, assessing and understanding the performance of multi-objective algorithms in software model refactoring is of paramount importance, especially when the goal is to integrate them into the design and evolution phases of software development.

In this paper, we present open challenges that, to the best of our knowledge, hinder the exploitation of search-based techniques within the context of quality attribute optimization of software architectures. 
We also describe the plan to overcome some of the listed open challenges.
\section{State of the art}\label{sec:related}

In the past ten years, approaches on software architecture multi-objective optimization have been developed to optimize various quality attributes (such as reliability and energy)~\cite{Martens:2010bn,5949650,DBLP:conf/qosa/MeedeniyaBAG10,10.1007/978-3-642-13821-8_8,CORTELLESSA2021106568}; with various degrees of freedom in the modification of architectures (such as service selection~\cite{Cardellini:2009:QRA:1595696.1595718}.

Recent research analyzes the capacity of two distinct multi-objective optimization algorithms to enhance non-functional features inside a particular architecture notation (\ie Palladio Component Model)~\cite{NI2021106565,10.1145/3132498.3132509,Becker:2009cl}.
The authors use architectural approaches to find the best solutions, which primarily include changing system parameters (such as hardware settings or operation requirements).

Menasce~\etal have provided a framework for architectural design and quality optimization, \cite{DBLP:conf/wosp/MenasceEGMS10}. This framework makes use of architectural patterns to help the search process (such as load balancing and fault tolerance).
The approach has two drawbacks: performance indices are computed using equation-based analytical models, which may be too simple to capture architectural details and resource contention; the architecture must be designed in a tool-specific notation rather than in a standard modeling language (as we do in this paper). 

A method for modeling and analyzing AADL architectures has been given by Aleti~\etal~\cite{DBLP:books/daglib/0030032}.
A tool that may be used to optimize various quality attributes while adjusting architecture deployment and component redundancy has also been introduced.

Cortellessa and Di Pompeo~\citep{CORTELLESSA2021106568} have presented a multi-objective framework aimed at improving the quality of architectural models specified by \aemilia~\citep{Arcelli:2018vo}. 
Cortellessa and Di Pompeo analyzed the sensibility of genetic algorithms when changing the configuration parameters.

Cortellessa~\etal~\citep{SEAA2021} have instead studied the impact of specific non-functional quality metric (\ie performance antipatterns~\citep{DBLP:journals/infsof/ArcelliCP18}) on the overall quality of Pareto frontiers.
In order to evaluate the overall quality of Pareto frontiers in this study, Cortellessa~\etal exploited established quality indicators within the search-based theory.

Di Pompeo and Tucci~\citep{SEAA2022} have investigated the effect of introducing a time budget to multi-objective optimization driven by non-functional quality attributes (such as performance and reliability) on the overall Pareto frontiers quality.
The idea beyond the approach is to introduce concepts of search-based techniques already investigated within different domains to the software model optimization context.
\section{Quality Attribute Optimization framework}\label{sec:framework}

\begin{figure}
    \centering
    \includegraphics[width=.9\linewidth]{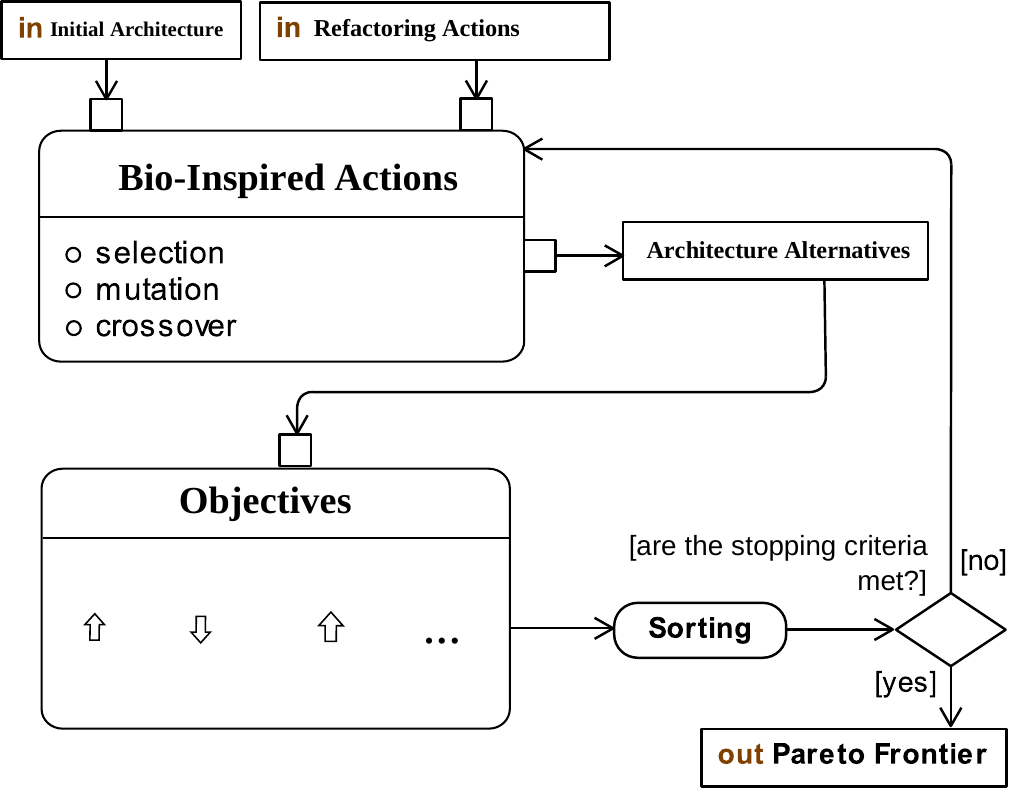}
    \caption{A typical framework for quality attributes optimization of software architecture.}
    \label{fig:framework}
\end{figure}

\Cref{fig:framework} depicts a classic search-based framework based on genetic algorithms. 
The framework starts from an \emph{Initial Architecture} and a set of \emph{Refactoring Actions}.
The initial architecture is the subject architecture to be optimized, while the refactoring actions are the set of all available actions that are combined by the \emph{Bio-Inspired Actions}, which are \emph{selection}, \emph{mutation}, and \emph{crossover}.

\noindent The \emph{crossover} operator mates solutions to evolve the species.
Several crossover policies, such as the single point crossover, can be employed in this step.
In this case, the chromosome will be split in two halves, and they will be alternatively combined.
While the crossover operation is performed, the \emph{mutation} operator randomly selects an element belonging to the chromosome and change it with a new one, as a genetic mutation would do in nature. The \emph{selection} operator is in charge of discarding the worst elements in the offspring (\ie the architecture alternatives).
Once the offspring is made, elements belonging to the offspring are sorted with respect to the objectives (\ie through the \emph{sorting} operator).
When there are more than one objective, we are in a multi-objective optimization process. Often, when the objectives are more than 3, we are in a many-objective optimization process.
Finally, when a stopping criterion is met, the optimization process ends and the final offspring will form the \emph{Pareto frontier}.
\section{Open Challenges}\label{sec:challenge}

\paragraph*{Lack of automation}
As introduced before, when search-based techniques are put in place they require to generate a higher number of alternatives, even thousands. 
Therefore, the automation to form alternatives is a must-have functionality in every model-based optimization framework.
Introducing automation in such a context would be a step ahead towards the adoption of search-based techniques within model-based optimization processes.
Furthermore, the automation is strictly related to the modelling notation and its expressiveness. 
To the best of our knowledge, we introduced the first refactoring engine for UML~\citep{DBLP:journals/infsof/ArcelliCP18}. Our refactoring engine exploits the Epsilon suite~\footnote{\url{https://www.eclipse.org/epsilon}}, and provides some facilities to refactor three different UML views, \ie Component, Sequence, and Deployment diagrams.
Then, we exploited the refactoring engine in a search-based optimization framework, where we sought optimal solutions with respect to four competitive objectives~\citep{SEAA2021, SEAA2022}.

\paragraph*{Problem formalization}
One of the relevant aspects for applying search-based techniques is the formalization of the problem. 
Often, search-based approaches exploit evolutionary algorithms and, in the majority of the cases, genetic algorithms are used to search the solution space.
Furthermore, a genetic algorithm is a bio-inspired algorithm that requires a chromosome to be manipulated for the evolution of the species.

To the best of our knowledge, there not exist guidelines in literature to represent specific problems as chromosomes. 
Positional structures are exploited to draw problems in model-based optimization studies~\citep{Koziolek:2011cg, NI2021106565}. 
Thus, each position of the chromosome have a fixed meaning. 
One of the advantages of using positional chromosomes is their fastest execution of bio-inspired actions on them.
However, the positional structure has the drawback to be too inexpressive within the context of model-based software refactoring.
To overcome the above limitation, there exists a chromosome representation that reports a refactoring action into a chromosome position~\citep{DBLP:journals/infsof/ArcelliCP18}.
Nevertheless, it is slower than the positional structure due to the complexity of compatibility checking among elements within chromosomes.

For the above issues,  we see an interesting research direction of the problem formalization within model-based software optimization that should fill the gap with more established search-based optimization problems.

\paragraph*{Time and resources requirements} 
Improving quality attributes by means of optimization techniques often requires a considerable amount of time and resources, as the search for better solutions relies on the manipulation of modelling artifacts.
Usually, the process of generating a new design alternative also involves a number of transformations from software design models (\eg UML) to non-functional models (\eg Queueing Networks, Markov Chains).
These target models are then used to quantify quality attributes like performance and reliability, either analytically or through simulation.
Given their inherent complexity and the toolchain employed in these contexts, it is generally challenging to make these activities more efficient.
As a consequence, they tend to extend the overall time needed for the optimization process.
Moreover, when this process is performed on models that are not just toy examples but realistic in size and complexity, it can last several days~\cite{SEAA2021}.
This clearly poses a challenge in adopting search-based optimization techniques in practical software engineering scenarios.
Finally, given the random nature of the algorithms that are usually employed in this context, it is very difficult to predict how long it will take for the process to complete.
This issue is exacerbated by the fact that, in most cases, the solution space is unknown to the designer at the beginning of the optimization.

\paragraph*{Architectural quality metrics}
When considering the multi-objective optimization in general, there is no lack of metrics (\eg quality indicators) that can be used to quantify the performance, and consequently the outcome, of the optimization process.
Nonetheless, such metrics only provide feedback that is based on the numerical values achieved by the solutions in the Pareto front for each objective.
This viewpoint is useful to quantify the improvement realized by the solutions in terms of quality attributes and with respect to the initial model.
However, the designer would not gain any feedback on how the architectural model itself changed during the process.
This makes it difficult to compare the solutions in the Pareto front with the initial architecture and among themselves.
Such a comparison is crucial because it guides the decision-making process of adopting a new design.
In this regard, quality metrics that represent architectural aspects like the change in the number of communication paths, in their length, in the complexity of components, or in the number of exchanged messages, could make this comparison practical, and avoid inspecting every solution to obtain enough knowledge to make an informed decision.

\paragraph*{Explainability}
As it is the case for many other optimization techniques, the solutions that are obtained through multi-objective optimization do not carry information about the specific causes that led to the generation and selection of such solutions.
For instance, at the end of an automated refactoring process guided by a genetic algorithm, while we can, of course, inspect the solutions to learn what refactoring actions were applied to obtain the best results, we have no knowledge about the circumstances that made those choices preferable to all the others.
In other words, we cannot explain why some modifications should be applied to our architecture other than for the quality attributes they seem to improve.
In order to understand the modifications, we would have to know why they are beneficial.
Unfortunately, in this kind of optimization processes, this is left for the designer to figure out.
In this sense, the lack of explainability of results makes it difficult for the designer to justify the new modifications that she is proposing on the basis of a completely automated process. 

\paragraph*{Reproducibility}
Reproducing results of optimization experiments has been an important concern in recent years, both for researchers and practitioners.
Of course, the main obstacle in this regard is represented by the random nature of many optimization techniques when it comes to the generation of new solutions and the exploration of the solution space.
On top of this, when optimizing architectures, the solution space is difficult or impossible to define beforehand.
More often than not, the solution space is not represented by just all the possible combinations of feasible modifications to the initial architecture, but it is built as the process goes on and the set of architectural elements that are possible targets of modifications changes.
This uncertainty in the definition of the solution space and, consequently, in the obtained results is usually tackled by performing multiple runs of the same experiments, and by trying to reach conclusions on the basis of the information gathered in all the runs.
While this is reasonable and practical in most cases, it can be very expensive when optimizing architectures, and less effective on large solution spaces.
Therefore, achieving perfect reproducibility is still a challenge in architectural optimization, and one that is rarely addressed by the relevant literature.

\section{Conclusion}\label{sec:conclusion}

In this paper, we reported open challenges of quality attributes optimization of software architectures.
To the best of our knowledge, these open challenges hinder the utilization of search-based techniques in software architecture optimization.

Furthermore, our agenda foster new research activities in the field of software architecture optimization, especially for the optimization of non-functional properties, such as performance and reliability.

As short-term future work, we will try to tackle the challenge about the \emph{lack of automation}.
We already presented approaches targeted at this challenge~\citep{SEAA2021, Arcelli:2018vo}.
However, we plan on extending the introduced automation by supporting more refactoring actions, for example by implementing the Fowler's refactoring portfolio~\citep{DBLP:conf/xpu/Fowler02}.

As mid-term future work, we will attempt to address the challenge about the \emph{architectural quality metrics}. 
We plan to exploit quality estimation techniques well-recognized in the search-based community~\citep{DBLP:journals/tse/LiCY22}.

As long-term future work, we will attempt to address the challenge about the \emph{problem formalization}.
To empirically address this challenge, a profound analysis of several software architectures is probably required, and each one will generate a specific optimization problem.

\section*{Acknowledgment}
Daniele Di Pompeo is supported by the Centre of EXcellence on Connected, Geo-Localized and Cybersecure Vehicle (EX-Emerge), funded by the Italian Government under CIPE resolution n. 70/2017 (Aug. 7, 2017).

Michele Tucci is supported by the OP RDE project No. CZ.02.2.69/\-0.0/\-0.0/\-18\_053/\-0016976 ``International mobility of research, technical and administrative staff at the Charles University''.

\bibliographystyle{IEEEtran}
\bibliography{IEEEabrv,biblio}

\end{document}